\documentclass[10pt,aps,pre,twocolumn,superscriptaddress,reprint]{revtex4-1}
\usepackage{graphicx}
\usepackage{float}
\usepackage{amsmath}
\usepackage{mathtools} 
\usepackage{color}
\usepackage{xr}
\usepackage[english]{babel}
\usepackage{silence}
\usepackage[T1]{fontenc}
\usepackage[utf8]{inputenc}

\begin{document}
\title{Memory formation in jammed hard spheres}
\date{\today}
\author{Patrick Charbonneau}
\affiliation{Department of Chemistry, Duke University, Durham, North Carolina 27708}
\affiliation{Department of Physics, Duke University, Durham, North Carolina 27708}
\author{Peter K. Morse}
\thanks{Corresponding author.}
\email{peter.k.morse@gmail.com}
\affiliation{Department of Chemistry, Duke University, Durham, North Carolina 27708}

\begin{abstract}
Liquids equilibrated below an onset density share similar inherent states, while above that density their inherent states markedly differ. Although this phenomenon was first reported in simulations over 20 years ago, the physical origin of this memory remains controversial. Its absence from mean-field descriptions, in particular, has long cast doubt on its thermodynamic relevance. Motivated by a recent theoretical proposal, we reassess the onset phenomenology in simulations using a fast hard sphere jamming algorithm and find it both thermodynamically and dimensionally robust. Remarkably, we also uncover a second type of memory associated with a Gardner-like change in behavior along the jamming algorithm.
\end{abstract}

\maketitle

The state of a material is nominally the product of its history, echoing both states and processes previously encountered. Yet equilibrium states are memoryless. Only certain non-equilibrium processes allow information to be stored, retained, and summoned back.  Because of the obvious uses for such memory, both nature and industry have developed myriad ways of harnessing it, including phase changes~\cite{ovshinsky_reversible_1968}, mechanical instabilities~\cite{pine_chaos_2005, mahadevan_self-organized_2005,bowden_spontaneous_1998,paulsen_multiple_2014,reid_auxetic_2018}, allostery~\cite{monod_nature_1965,rocks_designing_2017,yan_architecture_2017}, and wiping out~\cite{barker_magnetic_1983,sethna_hysteresis_1993}. Given their rich out-of-equilibrium physics, glass-forming materials exhibit all of these memory types, and thus broadly inform our understanding of them. Spin glass models, in particular, form the theoretical basis for both machine and biological learning~\cite{stein_spin_1992, choromanska_loss_2015, baity-jesi_comparing_2019}. Structural glasses, which are thought to be close relatives~\cite{parisi_theory_2020}, form an even richer array of memory types via out-of-equilibrium processes as varied as shearing~\cite{paulsen_multiple_2014, reid_auxetic_2018,keim_memory_2019}, heating cycles~\cite{debenedetti_supercooled_2001}, and aging~\cite{angell_relaxation_2000}.

Inherent state memory, which relates an equilibrium liquid state to its nearest energy minima or jammed configuration through fast out-of-equilibrium quenches~\cite{debenedetti_supercooled_2001,heuer_exploring_2008}, is one of the simplest types of memories in glasses.
What macroscopic properties of the original liquid can the inherent structure recall? In \emph{pure} $p$-spin models, which commonly inform the mean-field description of glasses \cite{charbonneau_glass_2017}, the answer is straightforward. Initial systems taken above the dynamical (or mode-coupling) transition temperature, $T_\mathrm{d}$, are quenched to inherent states indistinguishable from one another~\cite{cugliandolo_analytical_1993}. In other words, no information about the original liquid persits, other than that it was a liquid. This memorylessness has long been argued to be a general feature of glass-formers, but numerical simulations of (Kob-Andersen binary) Lennard-Jones liquids~\cite{sastry_signatures_1998,sastry_onset_2000,sastry_glass-forming_2013}, model polymers \cite{kamath_thermodynamic_2001}, and soft spheres~\cite{ozawa_jamming_2012,berthier_equilibrium_2016,jin_jamming_2020} do not concur. In these systems, all states prepared above an onset  $T_\mathrm{on}>T_\mathrm{d}$ share a same inherent state, but inherent states of liquids prepared below $T_\mathrm{on}$ differ. The resulting amorphous solid thus encode some features of the original liquid.

\begin{figure*}[ht]
\includegraphics[width=\linewidth]{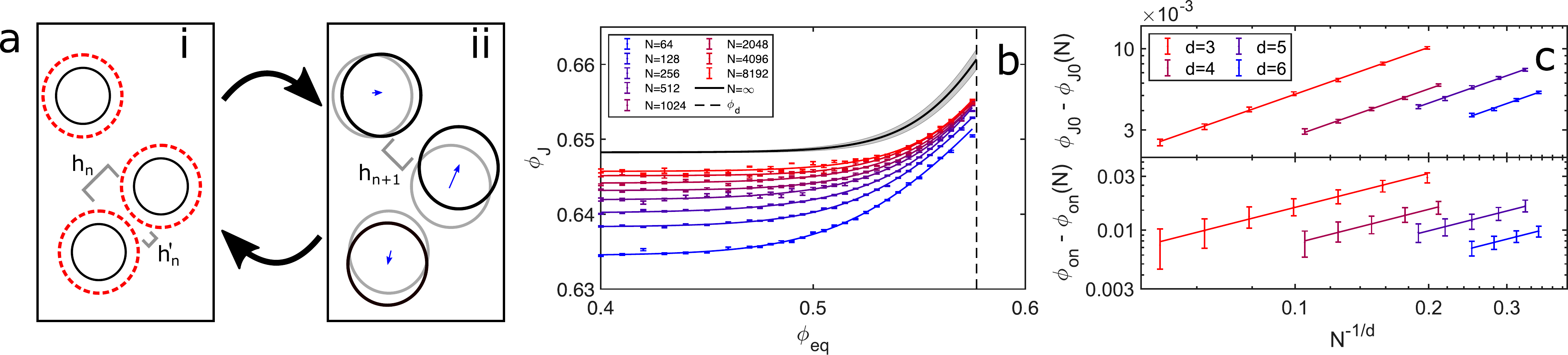}
\caption{\textbf{a)} Schematic of the two-step iterative jamming algorithm. \textbf{(i)} \textit{Inflation}: particles (black disks) separated by minimum gap $h_\mathrm{n}$ expand uniformly (red disks) until $h'_\mathrm{n}$. \textbf{(ii)} \textit{Repulsion}: an effective free energy is minimized until the minimum gap reaches $h_\mathrm{n+1} = h_\mathrm{n}$. The cycle is repeated until the density converges at jamming. \textbf{b)} The onset is clearly visible in $d=3$ for all system sizes considered. Lines are fits to the phenomenological crossover form Eq.~\eqref{eqn:onsetFit}. The thermodynamic $N\rightarrow\infty$ limit of the fit parameters (black line) shows that the onset appears well before the dynamical (mode-coupling) crossover (dashed black line). \textbf{c)} Finite-size scaling of the jamming transition $\phi_\mathrm{J0}$ below the onset (top), and the finite-size scaling of the onset  (below), where lines are fits to Eq.~\eqref{eqn:finSize}, and curves are offset for visual clarity.}
\label{fig:phiOnset}
\end{figure*}

Attempts to explain away this discrepancy abound. Finite-size \cite{crisanti_potential_2000,dasgupta_free-energy_2000} or finite-dimensional corrections \cite{dasgupta_free-energy_2000,glotzer_potential_2000} have been invoked, measurement protocols have been questioned~\cite{ozawa_jamming_2012,jin_jamming_2020}, as has the validity of the connection between spins and particle models~\cite{dasgupta_free-energy_2000}. The solution of the glass problem in the high-dimensional, $d\rightarrow\infty$ limit~\cite{parisi_theory_2020}, however, has revealed that the mean-field analogy between spins and particles glasses is quite strong, and some features of glass phenomenology are remarkably robust to dimensional changes~\cite{charbonneau_glass_2017}. 
The conceptual crisis was therefore complete when Folena et al.~\cite{folena_rethinking_2020,zamponi_surfing_2019} recently realized that \emph{mixed} $p$-spin models generically present an onset, and hence that \emph{pure} $p$-spin models might be exceptional rather than typical (see also~\cite{altieri_dynamical_2020}). 

While this advance offers a possible resolution of the original inconsistency, it does not address many of the remaining concerns, including algorithmic and finite-size considerations. In this letter, we use advanced computer simulations to eliminate these hypotheses and strongly evince the existence of a distinct landscape onset in liquids. We further uncover that the preparation algorithm itself has signatures of a dynamical transition, which can be used to define a memory that makes a distinction between all initial liquid conditions, even before the onset is reached.

\noindent \paragraph*{Model and Simulation Method---}
We consider the inherent states of hard sphere glass formers obtained by rapidly compressing, i.e., crunching, an equilibrated liquid at volume fraction $\phi_\mathrm{eq}$ to its nearest jamming point.
Existing crunching algorithms, however, either violate the hard sphere constraint~\cite{ohern_jamming_2003,ashwin_calculations_2012,morse_geometric_2014}, allow for significant equilibration~\cite{lubachevsky_geometric_1990,jin_jamming_2020}, or scale poorly with system size \cite{lerner_simulations_2013,rojas_inferring_2019}. In order to avoid these pitfalls, we modify a recent algorithm by Arceri and Corwin \cite{arceri_vibrational_2020} and propose an iterative two-step scheme based on the minimum scaled gap, ${h = \min_{ij}(h_{ij}) = \min_{ij}[d_{ij}/(r_i + r_j)]}$ between particles $i$ and $j$ of radii $r_i$ a distance $d_{ij}$ apart. Iterating step $n$ involves two sub-steps: \textit{inflation} and \textit{repulsion}. The former entails expanding particles uniformly, thus creating a new minimum gap, $h'_\mathrm{n} = \alpha h_\mathrm{n}$, and the latter uses the FIRE algorithm~\cite{bitzek_structural_2006} to minimize the effective thermal potential for hard spheres near jamming~\cite{brito_rigidity_2006, altieri_jamming_2016}, until $h_\mathrm{n+1} =h_\mathrm{n}$ (Appendix A). Although the minimal scaled gap stays constant from one step to the next, interparticle distances steadily decrease, and hence the algorithm converges at jamming. An expansion factor $\alpha<1$ ensures that the hard sphere constraint is never violated. Interestingly, a marked algorithmic slowdown of the FIRE minimization arises well before jamming is reached. 
We cap the number of steps of this minimization to a small multiple of the degrees of freedom, $n_\mathrm{FIRE}=\tau Nd$ to prevent a full minimization--and thus unwanted thermalization--as the crunching proceeds. Setting $\alpha=0.9$ and $\tau=2$, achieves the lowest jamming density while also creating a reliably rigid structure (Appendix B), thus ensuring that equilibration is maximally suppressed along the process. A low-density fluid thus crunched hence best approximates the maximally random jammed state~\cite{torquato_is_2000}. 

\begin{figure}[ht]
\includegraphics[width=\columnwidth]{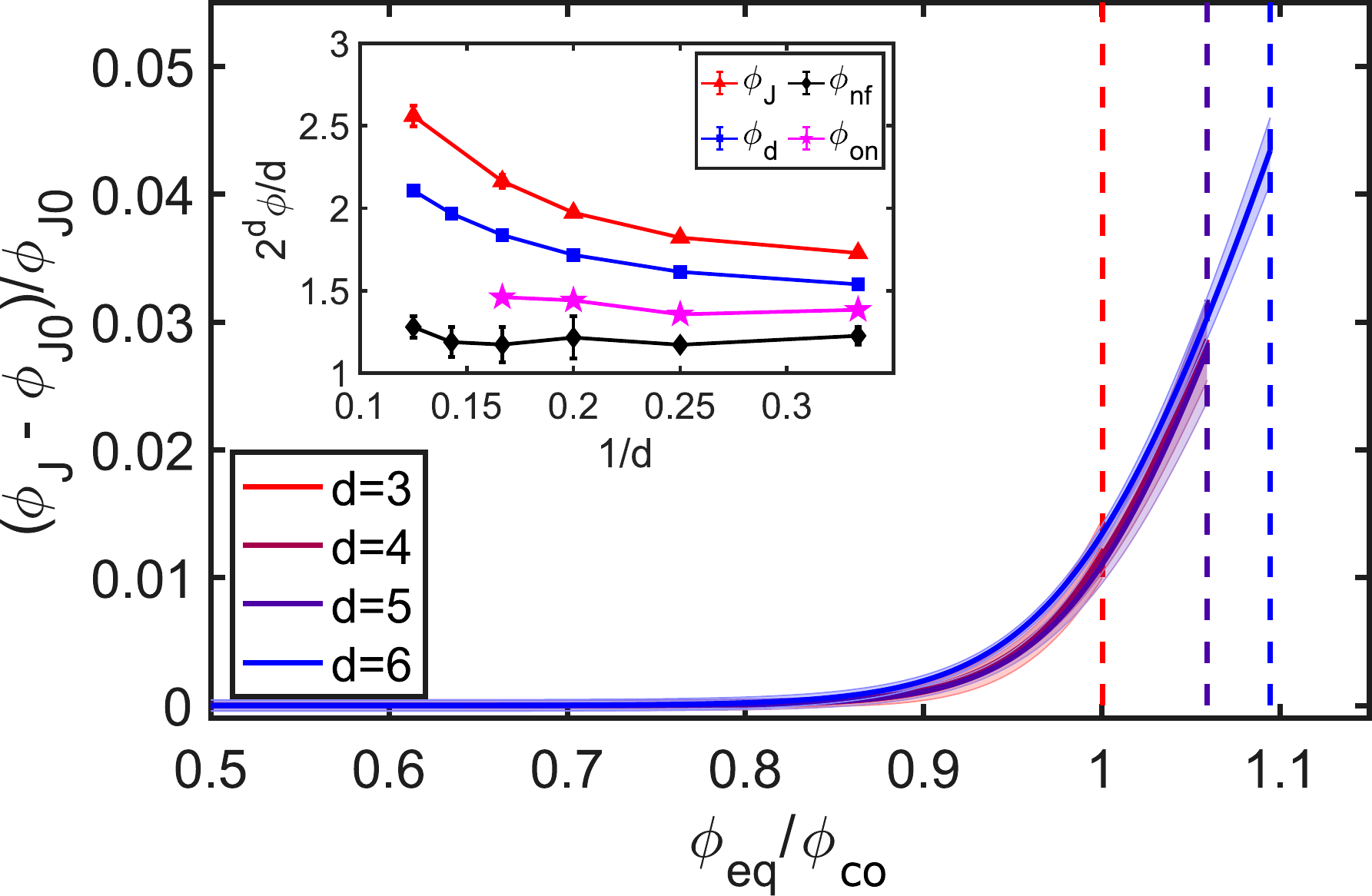}
\caption{Infinite-system size onset curves for different dimensions can be collapsed, suggesting that the inherent structure onset exists in both the thermodynamic and the infinite-dimensional limits.  Shaded regions give the standard error of Eq.~\eqref{eqn:onsetFit} with $95\%$ confidence intervals on parameters and dashed lines denote $\phi_\mathrm{d}$ from Ref.~\cite{charbonneau_hopping_2014}. The steady increase of $\phi_\mathrm{d}$ with dimension on this scale shows that $\phi_\mathrm{co}<\phi_\mathrm{d}$. The collapse further suggests the indentification $\phi_\mathrm{on}\sim 0.9\phi_\mathrm{co}$. \textbf{(Inset)} Scaling of $\phi_\mathrm{J0}$ and $\phi_\mathrm{on}$ with $d$, compared with those for the avoided dynamical transition, $\phi_\mathrm{d}$, and the onset of non-Fickian diffusion, $\phi_\mathrm{nf}$, from Ref.~\cite{charbonneau_hopping_2014} reveals that both $\phi_\mathrm{nf}$ and $\phi_\mathrm{on}$ exhibit a trivial mean-field-like dimensional scaling down to physical dimensions, unlike that of $\phi_{\mathrm{d}}$ and $\phi_{\mathrm{J0}}$.}
\label{fig:finSizeScaling}
\end{figure}

\paragraph*{Onset Memory---}
The first quantity of interest is the density of jammed states $\phi_\mathrm{J0}(N)$, obtained from low-density liquids, 
and its scaling with system size $N$ upon approaching the thermodynamic $N\rightarrow\infty$ limit.
Because of the critical nature of jamming, we expect
\begin{equation}
\phi_\mathrm{J0} - \phi_\mathrm{J0}(N) \sim N^{-1/\nu d} 
\label{eqn:finSize}
\end{equation}
with correlation length exponent $\nu$. Soft spheres studies have found $\nu \approx 0.7$~\cite{ohern_random_2002,ohern_jamming_2003, ozawa_jamming_2012}, which is inconsistent with $\nu\approx1$ obtained from direct measurements of the correlation length at jamming~\cite{vagberg_finite-size_2011}. We here robustly find $\nu\approx 1$ in all $d$, with $\nu = 1.01 \pm 0.04$, $0.99 \pm 0.06$, $1.01 \pm 0.10$, and $1.0 \pm 0.3$ in $d=3$, $4$, $5$, and $6$ respectively, thus resolving the discrepancy.  Although different exponents can in principle be attributed to model and algorithmic differences \cite{vagberg_finite-size_2011}, the scaling difference between soft and hard spheres might also originate from the fact that minimization of the former, unlike crunching of the latter, allows for weak barriers to be crossed. In support of this hypothesis, we note that our thermodynamic extrapolations for $\phi_\mathrm{J0}$ are close to but systematically smaller than those for soft spheres for all dimensions considered (Appendix C), including the careful estimate of Ref.~\onlinecite{ohern_random_2002}. In addition, the lack of dimensional dependence of this particular critical exponent for a specific model and algorithm gives further credence to $d_u=2$ being lower critical dimension for jamming~\cite{wyart_rigidity_2005,goodrich_finite-size_2012,hexner_can_2019}. 
 
Figure \ref{fig:phiOnset}c shows a clear dependence of the inherent state on the original equilibrium liquid condition, such that for $\phi_\mathrm{eq}\lesssim\phi_\mathrm{on}$, $\phi_\mathrm{J}$ is constant, and for $\phi_\mathrm{eq}\gtrsim\phi_\mathrm{on}$, $\phi_\mathrm{J}$ increases with $\phi_\mathrm{eq}$. The change from one regime to the other, however, does not sharpen as the system size increases, and thus remains a crossover in the thermodynamic limit.
To quantify this feature, we use the empirical softmax form \cite{dugas_incorporating_2001}

\begin{equation}
\phi_\mathrm{J}(\phi_\mathrm{eq}) = \phi_\mathrm{J0} + ab\ln (1 + e^{(\phi_\mathrm{eq}-\phi_\mathrm{co})/b}),
\label{eqn:onsetFit}
\end{equation}

\noindent where $\phi_\mathrm{co}(N)$ marks the crossover point between the low density and high density linear regimes, $a = \frac{d\phi_\mathrm{J0}}{d\phi_\mathrm{eq}}$ for $\phi_\mathrm{eq} \gg \phi_\mathrm{co}$, and $b(N)$ characterizes the width of the crossover region. This form nicely recapitulates our observations, but we note that $\phi_\mathrm{co}$ occurs well above the point at which $\phi_\mathrm{J}$ deviates from $\phi_\mathrm{J0}$, which traditionally defines the onset. Without loss of generality, we thus define $\phi_\mathrm{on} = 0.9 \phi_\mathrm{co}$. The result scales as $\phi_\mathrm{on} \sim N^{-1/d}$ (Figure \ref{fig:phiOnset}b). Because of the limited density range between $\phi_\mathrm{co}$ and $\phi_\mathrm{d}$, around which standard computations become particularly onerous for monodisperse systems, the fitting parameters $a$ and $b$, cannot be independently determined at fixed $N$. Imposing that a single $a$ should fit all $N$, however, suffices to obtain a robust extrapolation of Eq.~\eqref{eqn:onsetFit} to the thermodynamic limit (Appendix D).

In order to compare the dimensional trend quantitatively, we consider the fractional deviation from $\phi_\mathrm{J0}$ with the normalized density growth $\phi_\mathrm{eq}/\phi_\mathrm{co}$. The thermodynamic onset results then collapse onto a master curve (Fig.~\ref{fig:finSizeScaling}), strongly suggesting that the onset persists as a crossover as $d\rightarrow\infty$. This scaling also shows that $\phi_\mathrm{co}$ and thus $\phi_\mathrm{on}$ are numerically distinct from the (avoided) dynamical transition $\phi_\mathrm{d}$ as indicated by the steady increase of $\phi_\mathrm{d}$ on this scale. 
Hence, independently of the proposed scaling, our results validate earlier numerical studeis and are in sharp contrast with those of Ref.~\cite{cugliandolo_analytical_1993} for pure $p$-spin models.  The inset of Fig.~\ref{fig:finSizeScaling} suggests that upon considering the mean-field, $d\rightarrow\infty$, limit the onset remains roughly constant, while the (avoided) dynamical transition shifts markedly as $d$ increases.  Interestingly, this same qualitative behavior has been observed for another onset, that of non-Fickian diffusion, $\phi_\mathrm{nf}$~\cite{charbonneau_hopping_2014}.

From a theoretical standpoint, these various results are quite informative. While the (avoided) dynamical transition of liquids is sensitive to structure (especially compared to that of structureless liquids~\cite{charbonneau_hopping_2014,mangeat_quantitative_2016}), both $\phi_\mathrm{on}$ and $\phi_\mathrm{nf}$ are not. This distinction suggests that separate underlying (landscape) mechanisms underlie the two types of features. Although it is not immediately apparent why  $\phi_\mathrm{on}$ and $\phi_\mathrm{nf}$ should scale similarly, the robustness of our results suggests that a complete out-of-equilibrium dynamical theory should account for their (near) coincidence.

\begin{figure}[htb]
\includegraphics[width=\linewidth]{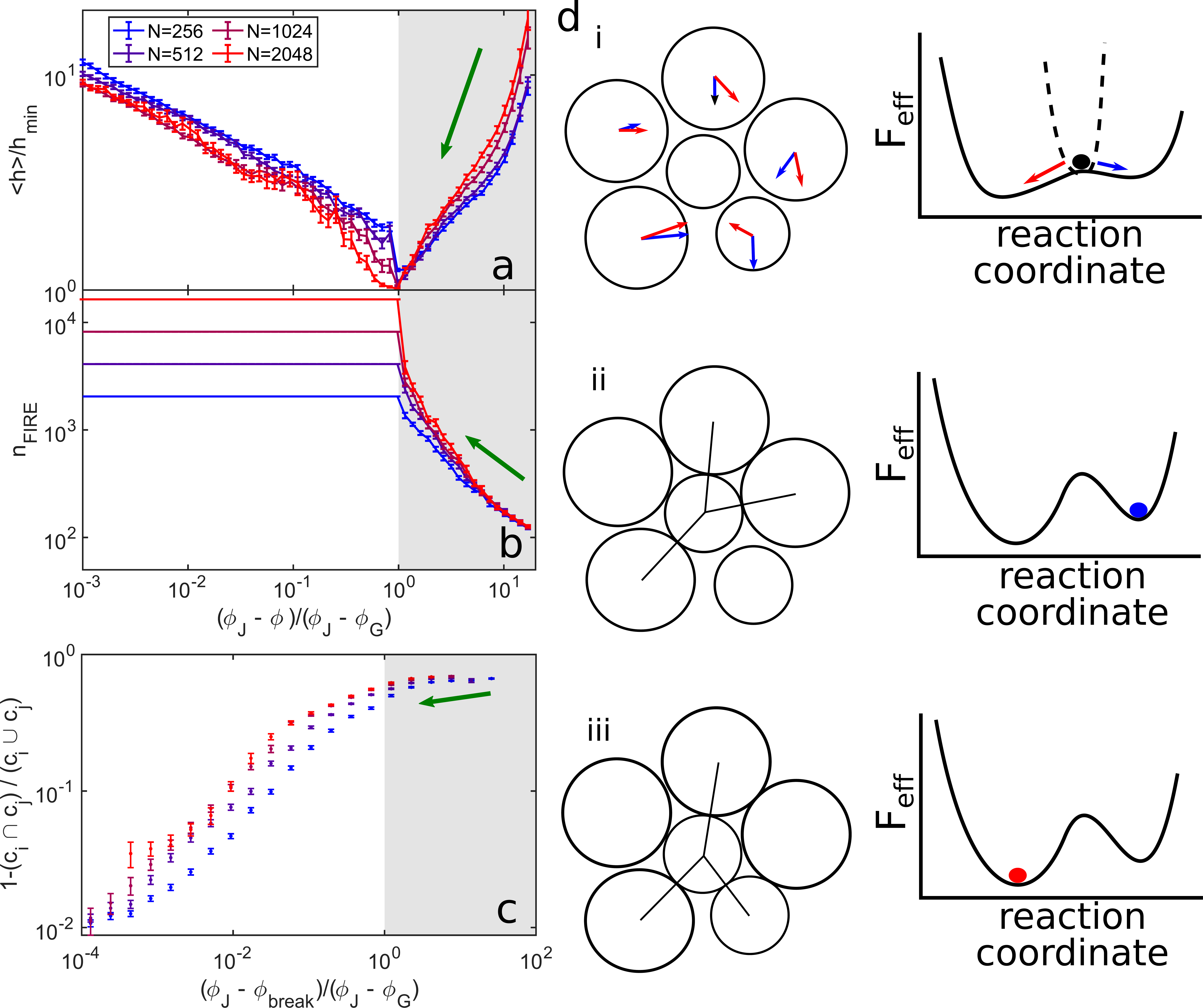}
\caption{The onset of the algorithmic slowdown at $\phi_\mathrm{G}$ is simultaneously characterized by three observables, which robustly identify a change in the crunching process. (Results for $d=4$ with $\phi_\mathrm{eq} = 0.2$, which are shown here, are typical of other dimensions and conditions.) \textbf{a)} At $\phi_\mathrm{G}$,  the distribution of gaps narrows significantly, such that the minimum gap most closely approaches the average gap. \textbf{b)} The number of minimization loops necessary to complete the repulsion sub-step (ii) of the jamming algorithm grows precipitously, and is manually cut off at $\tau=2$. 
\textbf{c)} Comparing the contact network, $c_i$, for an unperturbed system at jamming with the network, $c_j$, for a replica perturbed at $\phi_\mathrm{break}$ shows that systems perturbed before $\phi_\mathrm{G}$ (gray zone) end up with markedly different contact networks while systems perturbed beyond $\phi_\mathrm{G}$ (white zone) exhibit growing similarities. Increasing system size makes the effect more prominent and shifts the process to higher densities that nonetheless remain distinct from $\phi_\mathrm{J}$ (Appendix E). \textbf{d)} Taken together, these observation suggest that saddles start to dominate the landscape of the crunch algorithm around $\phi_\mathrm{G}$, thus resulting in sluggish dynamics and a large contact network response to small perturbations in particle positions. In other words, a slightly perturbed system (i) then jams as (iii), whereas the original system jams at (ii). This series of observations for an out-of-equilibrium algorithm is reminiscent of the Gardner-like behavior of quasi-static state followings in ultrastable glasses~\cite{berthier_growing_2016}}
\label{fig:gardnerDef}
\end{figure}

\paragraph*{Algorithmic Memory---}
Surprisingly, a second form of memory develops before jamming is reached. As a liquid is initially crunched, interparticle gaps first grow more regular, such that  $\langle h \rangle/h_\mathrm{min} \sim 1$ (Figure \ref{fig:gardnerDef}a). Because of the disordered, and thus frustrated, nature of the jammed state, however, the repulsion sub-step becomes increasingly computationally arduous, as illustrated by the rapid growth in the number of minimization loops necessary to achieve $h_\mathrm{n+1} = h_\mathrm{n}$ (Fig.~\ref{fig:gardnerDef}b). Gap regularization then also goes into reverse. Remarkably, the two phenomena coincide at some $\phi_\mathrm{G}$. This putative algorithmic onset can be further characterized by considering the result of perturbing a state along the jamming algorithm. Taking exact replicas at $\phi_\mathrm{break}$ and applying a single Metropolis Monte Carlo step  before crunching anew gives rise to force contacts at jamming, $c_i$, that can vary. 
Comparing these contact networks using ${1-(c_i \cap c_j)/(c_i \cup c_j)}$, in particular, highlights structural differences. The quantity vanishes if the packings are identical and unity if the packings share no contacts. Figure \ref{fig:gardnerDef}c indicates that applying a perturbation before $\phi_\mathrm{G}$ results in markedly different jammed states, whereas perturbations made after $\phi_\mathrm{G}$ present increasingly small deviations (Appendix E).

Taken together these observations suggest that saddles start to dominate the optimization landscape around $\phi_\mathrm{G}$, forcing the selection of a nearby sub-basin and thus of a contact network at jamming (Fig.~\ref{fig:gardnerDef}d). A transition which sharpened with system size above $\phi_\mathrm{G}$ would imply that all replicas perturbed after $\phi_\mathrm{G}$ converge on the same contact network. That it does not suggests instead a rich, multi-layered landscape structure reminiscent of an equilibrium Gardner transition~\cite{charbonneau_glassy_2019, charbonneau_glass_2017, berthier_growing_2016, charbonneau_fractal_2014}, for which mean-field theory predicts a fractal hierarchy of sub-basins \cite{charbonneau_fractal_2014}.

\begin{figure}[ht]
\includegraphics[width=\linewidth]{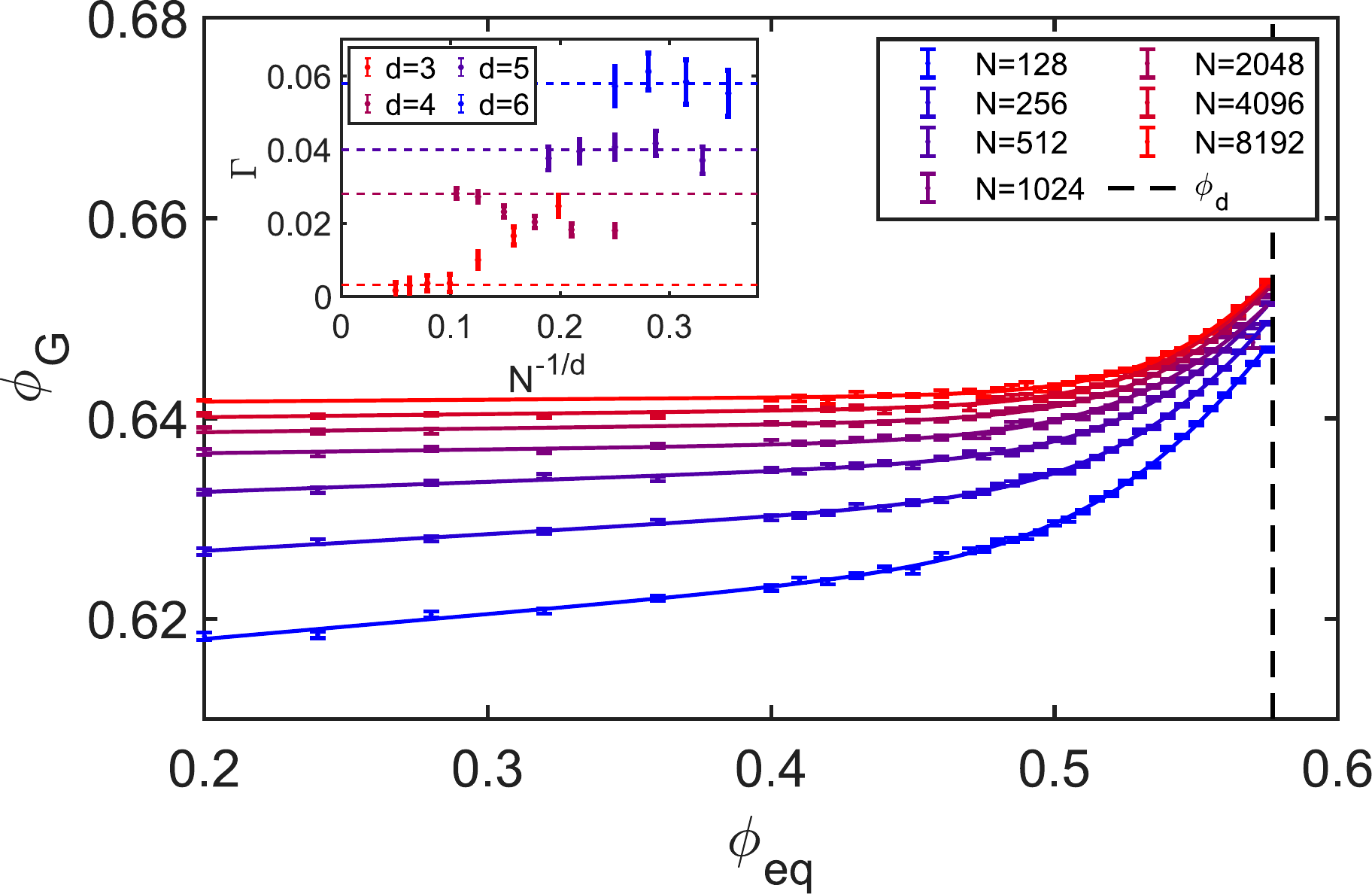}
\caption{\textbf{a)} The algorithmic $\phi_\mathrm{G}$ in $d=3$, identified as in Fig.~\ref{fig:gardnerDef}, shifts with system size. \textbf{(Inset)} The low-density slope of $\phi_\mathrm{G}$ tends to a finite value as $N$ increase in all dimensions (dashed lines). Because the density dependence of $\phi<\phi_\mathrm{on}$ seemingly persists in the thermodynamic limit, memory of the initial state appears upon crunching.}
\label{fig:gardnerSlope}
\end{figure}

The evolution of $\phi_\mathrm{G}$ upon increasing $\phi_\mathrm{eq}$ is akin to that of $\phi_\mathrm{J}$ (Fig.~\ref{fig:gardnerSlope} and Appendix D) but with an initial linear growth instead in lieu of a density-independent regime. To estimate if both this linear scaling and $\phi_\mathrm{G}$ persist with increasing system size and dimension, we fit the results to a modified form of the softmax potential 

\begin{equation}
\phi_\mathrm{G}(\phi_\mathrm{eq}) = \phi_\mathrm{G0} + \Gamma(\phi_\mathrm{eq}-\phi_\mathrm{co}) + (a-\Gamma)b\ln (1 + e^{(\phi_\mathrm{eq}-\phi_\mathrm{co})/b}),
\label{eqn:onsetFitG}
\end{equation}

\noindent where $a$, $b$, and $\phi_\mathrm{co}$ are taken from fits to Eq.~\eqref{eqn:onsetFitG}, and ${\Gamma = \left.\frac{d\phi_\mathrm{G}}{d\phi_\mathrm{eq}}\right|_{0}}$ is the slope of the linear regime.  Figure~\ref{fig:gardnerSlope} shows that $\Gamma$ tends to a constant as $N\rightarrow\infty$, and that this constant increases as $d$ increases (see also Appendix D). Hence, although systems prepared at different $\phi^{(1)}_\mathrm{eq}<\phi^{(2)}_\mathrm{eq}<\phi_\mathrm{on}$ both jam at a same density $\phi_\mathrm{J0}$, $\phi^{(1)}_\mathrm{eq}$ encounters a saddle-dominated regime at smaller densities than $\phi^{(2)}_{\mathrm{eq}}$. In other words, while the jammed state may not recall the liquid density used to prepare it, its crunching does.

The identification of $\phi_\mathrm{G}$, its similarity to a Gardner transition, and its signature of the onset provide guidance for solving out-of-equilibrium dynamical theories~\cite{agoritsas_out--equilibrium_2018,agoritsas_out--equilibrium_2019,altieri_dynamical_2020}. Indeed, while quasi-equilibrium calculations find that a Gardner transition is a necessary step towards jamming for liquids equilibrated beyond $\varphi_{\mathrm{d}}$\cite{berthier_growing_2016,berthier_gardner_2019}, our results suggest that an equivalent out-of-equilibrium phenomenon should be uncovered in a mean-field description. If true, this would resolve the paradoxical observation that jamming criticality is obtained in experimentally relevant regime \cite{ohern_jamming_2003,goodrich_finite-size_2012,lerner_low-energy_2013,charbonneau_jamming_2015}, with $\phi \ll \phi_\mathrm{on}$, even in the absence of standard Gardner physics. 

\paragraph*{Conclusion---}
By devising an efficient crunching algorithm that does not violate the hard-sphere condition, we have determined that inherent state memory persists in the thermodynamic and high-dimensional limits. 
Such memory thus ought to exist in mean-field descriptions. We have further identified a Gardner-like point in the strongly out-of-equilibrium behavior of our crunching algorithm. This quantity itself varies across $\phi_\mathrm{eq}$, and thus recalls the original liquid, even at densities well below the inherent structure onset. Although the location of this phenomenon is likely strongly algorithm dependent, we expect all such procedures to encounter a comparable slowdown or instability. Revisiting such algorithms might be particularly instructive, and could offer a lens to broader class of problems, particularly within generalized learning algorithms, for which interest in Gardner physics has recently grown \cite{abbara_rademacher_2020}. If the association is confirmed, then experimental validations of the Gardner physics would then also be well within reach.

\begin{acknowledgments}
We acknowledge many stimulating discussions with Ada Altieri, Francesco Arceri, Silvio Franz, Jorge Kurchan, Giorgio Parisi, and Francesco Zamponi. This work was supported by the Simons Foundation grant \# 454937. Most simulations were performed at Duke Compute Cluster (DCC), for which the authors thank Tom Milledge's assistance. The authors also thank the Extreme Science and Engineering Discovery Environment (XSEDE), supported by National Science Foundation Grant No. ACI-1548562, for computer time. Data relevant to this work have been archived and can be accessed at the Duke Digital Repository \url{https://doi.org/10.7924/XXXXXXX}.
\end{acknowledgments}




\appendix

\section{Logarithmic Potential}

The effective potential used in the repulsion step of our jamming algorithm is given by $U = \sum_{i>j} U_{ij}(h_{ij})$ with

\begin{equation}
U_{ij}(h_{ij}) = \begin{cases} 
\infty& h_{ij} < 0\\
-\ln(h_{ij}) + \frac{1}{c}(1+h_{ij})& 0 < h_{ij} < c \\
0& h_{ij} \ge c
\end{cases}
\end{equation}

\noindent where $h_{ij} = \frac{d_{ij}}{r_i + r_j}-1$ is the scaled gap between particles $i$ and $j$ with radii $r_i$ and $r_j$ and interparticle distance $d_{ij}$. Particles are monodisperse, except in $d=3$, where they are bidisperse with a 50:50 mixture of size ratio 1:1.4. The cutoff $c$ is set such that there is an average of $2d$ contacts per particle with $U_{ij}(h_{ij}) > 0 $. This ensures that the potential smoothly goes to zero when particles lose contact and allows only nearby particles to influence one another.

\section{Algorithm optimization and jamming criterion}

\begin{figure*}[htb]
\includegraphics[width=0.8\linewidth]{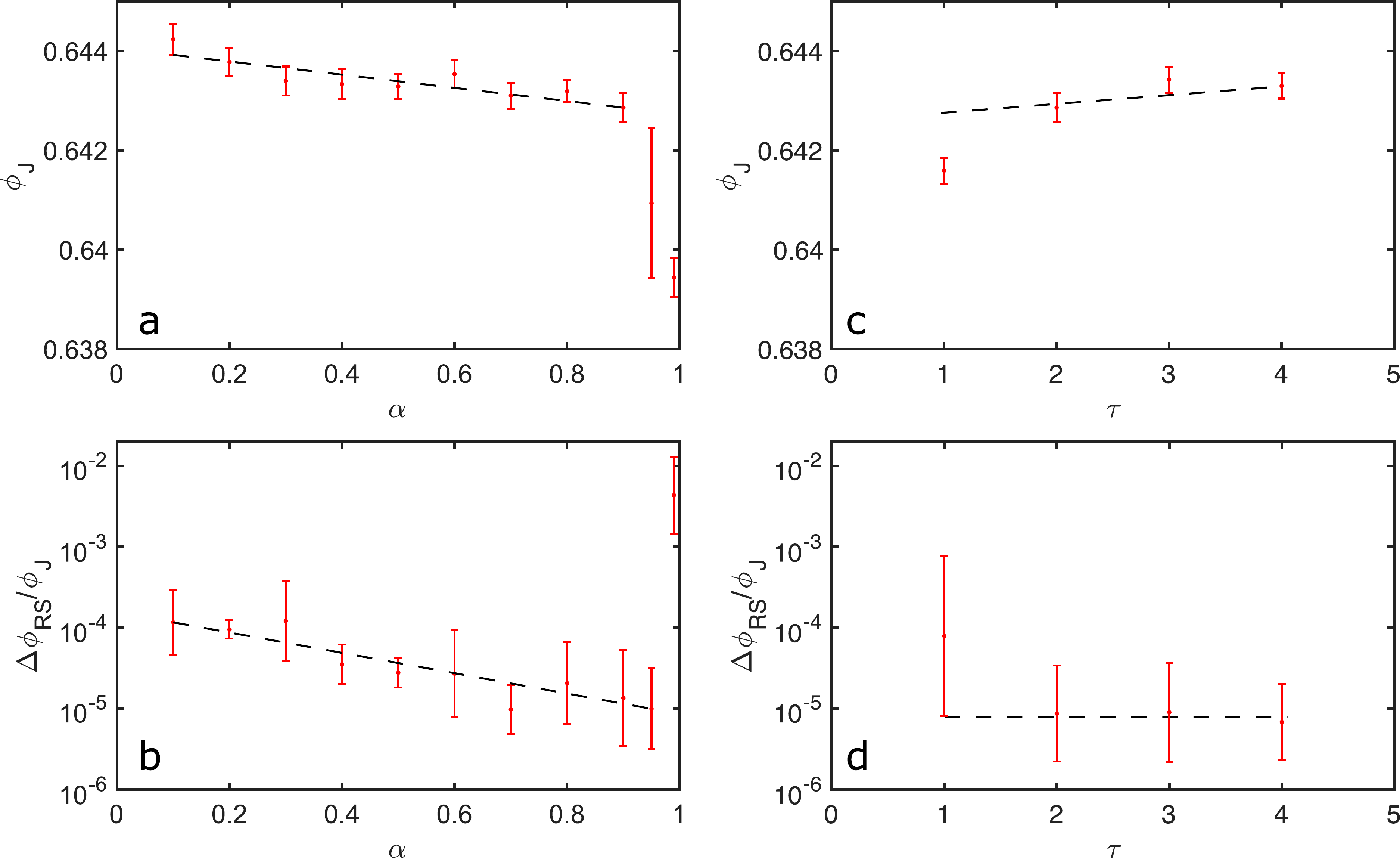}
\caption{Determination of the optimal parameters $\alpha$ and $\tau$ in the crunching algorithm for (generic conditions) $d=3$, $N=1024$, and $\phi_\mathrm{eq} = 0.4$. Optimal $\alpha$ and $\tau$ should result in: (i) minimal inflation necessary to jam the corresponding soft sphere system $\phi_{RS}$, and (ii) a minimal jamming density. \textbf{a)} $\langle \phi_\mathrm{J0} \rangle$ steadily decreases with increasing $\alpha$ at fixed $\tau=2$, hence a larger $\alpha$ results in better inherent states. \textbf{b)} $\phi_\mathrm{RS}$ decreases steadily, but rises sharply for $\alpha > 0.95$. \textbf{c)} $\phi_\mathrm{J}$  increases monotonically with $\tau$ at fixed $\alpha=0.9$, hence a smaller $\tau$ results in better inherent states. \textbf{d)} $\phi_\mathrm{RS}$ is constant above $\tau = 2$, but raises significantly for $\tau=1$. The combination $\alpha=0.9$ and $\tau=2$ is therefore near optimal per our selection criteria. Dashed lines are provided as trend guides.}
\label{fig:paramSweep}
\end{figure*}

The crunching algorithm described in the text requires that two parameters be optimized: the expansion parameter $\alpha$, which sets how much gaps shrink in the expansion substep, and the stopping parameter $\tau = 2$, which indicates the maximum number of FIRE minimization steps in the repulsion substep. In general, $\phi_\mathrm{J}$ decreases with increasing $\alpha$ and decreasing $\tau$, but extreme values result in mechanically unstable packings. We thus aim for parameters that minimize the inherent state density, $\phi_\mathrm{J}$, subject to the algorithm producing a stable jammed packing. 

First, each crunch is run until the relative density difference between steps $n$ and $n+1$ is $\frac{\phi_\mathrm{n+1}}{\phi_\mathrm{n}} - 1 < 10^{-10}$, which is set low enough to ensure that $\phi_\mathrm{J}$ does not further evolve. Crunching beyond this point is possible, but numerically wasteful for our needs. The quality of the resulting packing is then assessed by measuring the distance from a putative mechanically stable state. This distance is determined by inflating spheres (allowing overlaps) without minimization until the system has at least one excess contact above isostaticity, after removing rattlers, $\Delta\phi_\mathrm{RS}$ (found via a binary search algorithm)\cite{goodrich_finite-size_2012}. We finally choose $\alpha$ and $\tau$ so as to produce the smallest $\phi_\mathrm{J}$ possible without significantly increasing $\Delta\phi_\mathrm{RS}$. Figure~\ref{fig:paramSweep}b shows a jump in $\Delta \phi_\mathrm{RS}/\phi_\mathrm{J}$ for $\alpha > 0.95$ indicating that $\alpha > 0.95$ produces non-rigid packings and implying an optimum near $\alpha = 0.95$. Because the variance in $\phi_\mathrm{J}$ is then large, however, we conservatively set $\alpha = 0.9$. Similarly, Fig.~\ref{fig:paramSweep}e shows a rapid increase in $\Delta\phi_\mathrm{RS} / \phi_\mathrm{J}$ for $\tau < 2$ leading to our choosing $\tau = 2$.

\section{Comparison to soft sphere jamming}

\begin{figure*}[htb]
\includegraphics[width=0.8\linewidth]{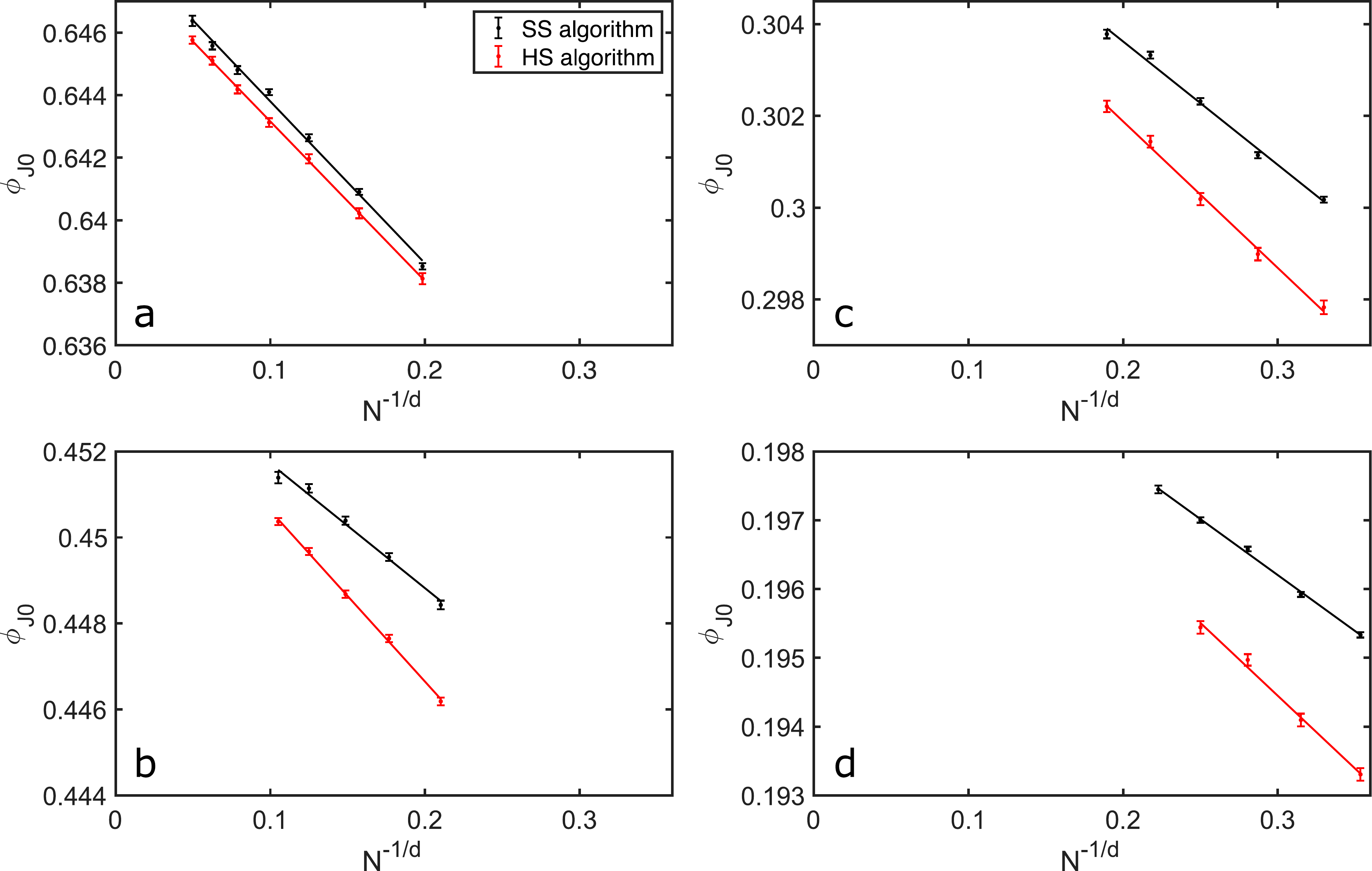}
\caption{Comparison of $\phi_\mathrm{J0}$ between the HS algorithm used in the text, and the SS protocol described in Ref.~\cite{morse_geometric_2014} for \textbf{a)} $d=3$, \textbf{b)} 4, \textbf{c)} 5, and \textbf{d)} 6,. The slopes appear consistent between algorithms, indicating a similar value of $\nu$, but the asymptotic value of $\phi_\mathrm{J0}$ reached by hard sphere crunching is significantly lower than for its soft sphere counterpart. Lines are best fits to Eq.~\eqref{eqn:finSize}.}
\label{fig:HSVsSS}
\end{figure*}

\begin{figure*}[htb]
\includegraphics[width=0.8\linewidth]{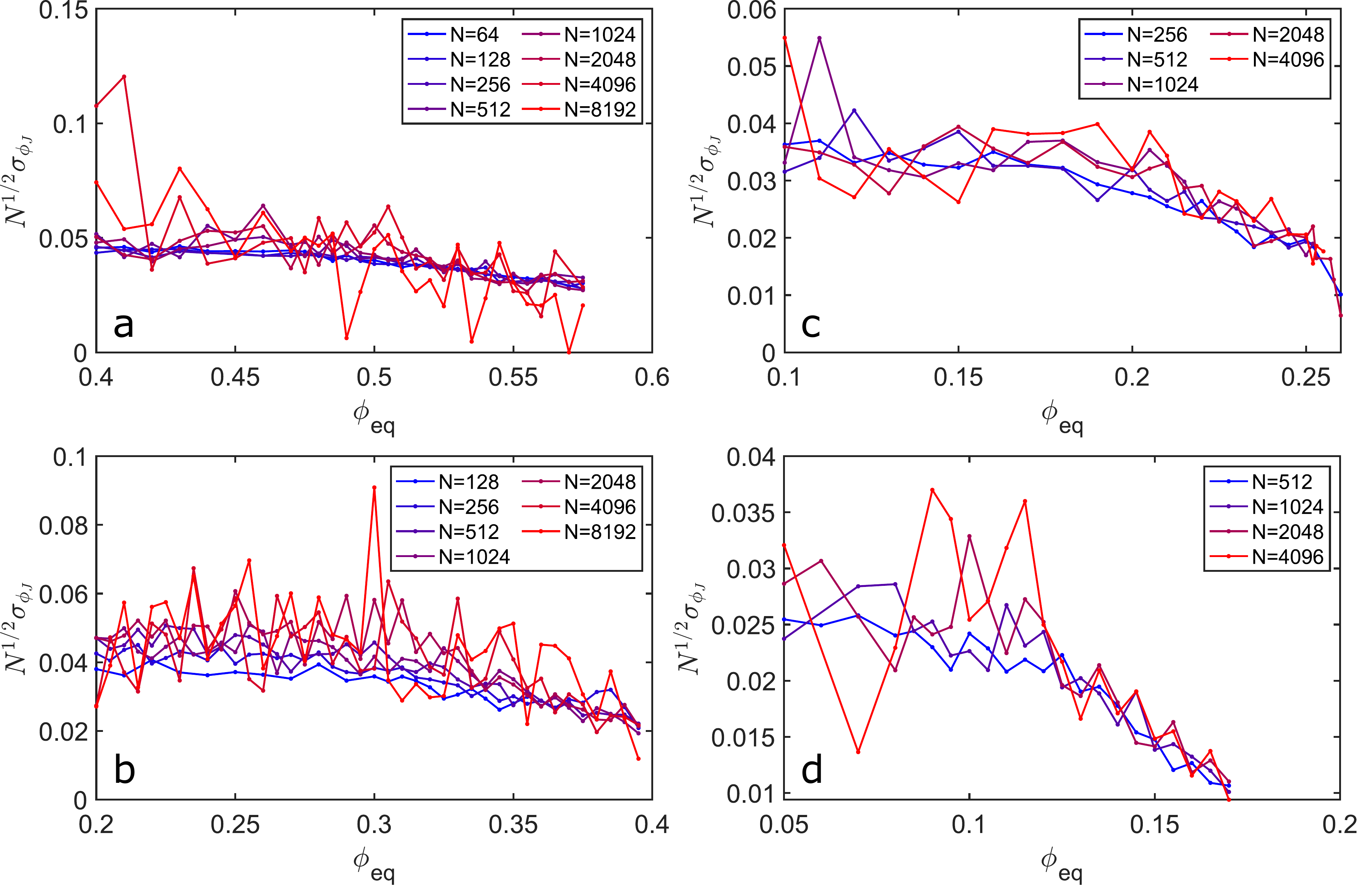}
\caption{The variance of the jamming density distribution, scaled as $N^{-1/2}$, collapses results for all \textbf{a)} $d=3$, \textbf{b)} $d=4$, \textbf{c)} $d=5$, and \textbf{d)} $d=6$. The spread of available jammed states decreases decreases with increasing $\phi_\mathrm{eq}$.}
\label{fig:phiJwidth}
\end{figure*}

The jamming density of hard sphere (HS) crunching algorithm produces lower inherent state densities than the infinite-temperature quench soft sphere (SS) geometric-mean-search algorithm described in Ref.~\cite{morse_geometric_2014}, as demonstrated in Fig. \ref{fig:HSVsSS}. 
The density spread of jammed densities, $\sigma_{\phi_\mathrm{J}}$, scales as 

\begin{equation}
\sigma_{\phi_\mathrm{J}} \sim N^{-1/2}
\label{eqn:phiJfluct}
\end{equation}

\noindent for all dimensions and $\phi_\mathrm{eq}$, consistent with results from earlier algorithms, e.g., \cite{ohern_jamming_2003} (Fig.~\ref{fig:phiJwidth}).

\section{High-dimensional scaling and fit parameters}

\begin{figure*}[htb]
\includegraphics[width=0.9\linewidth]{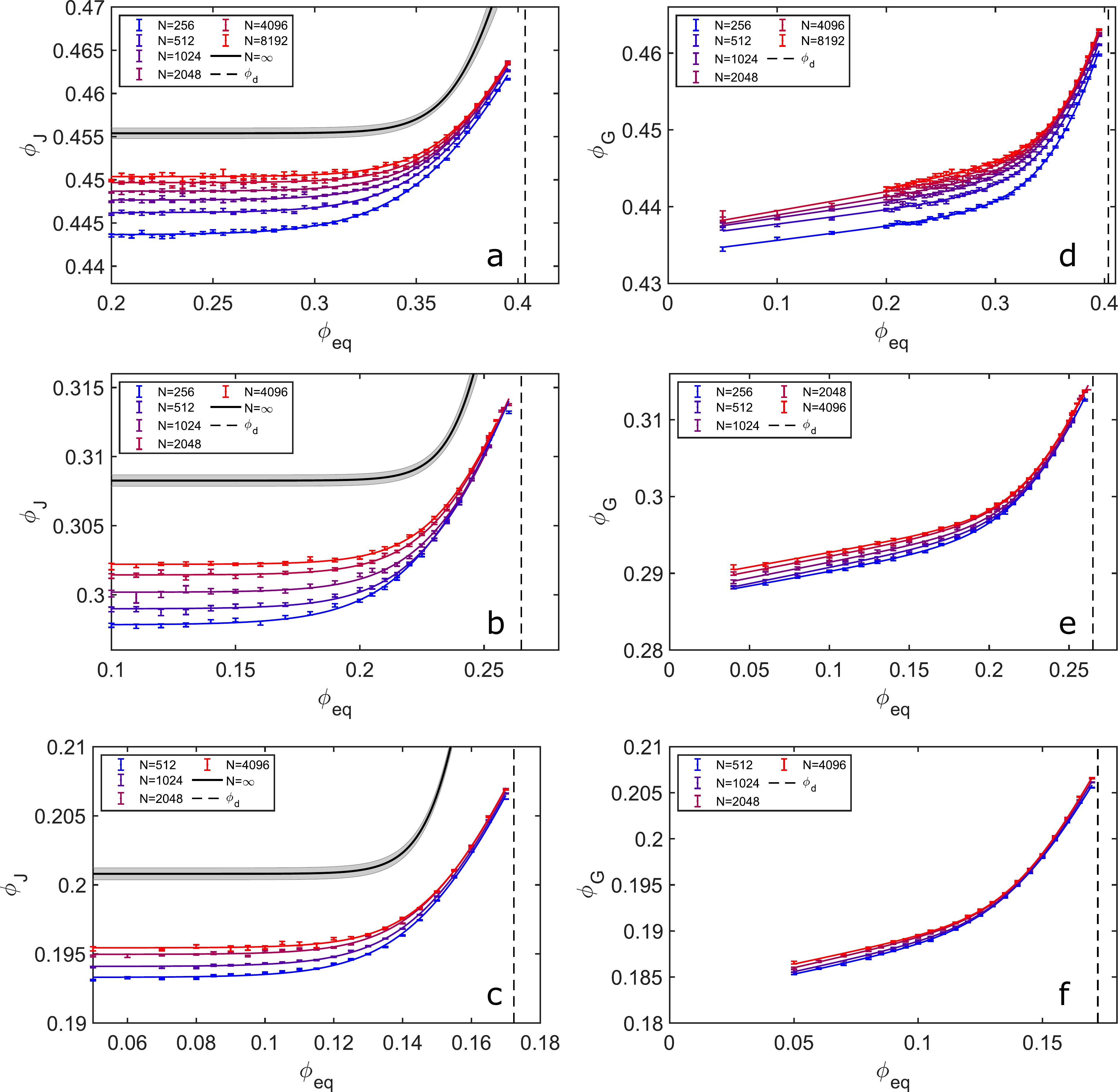}
\caption{The signature of the onset demonstrated in Fig.~1b is given for \textbf{a)} $d=4$, \textbf{b)} $5$, and \textbf{c)} $6$. Lines are fits to Eq.~\eqref{eqn:onsetFit}. In each case, the thermodynamic limit extrapolated from Eq.~\eqref{eqn:onsetFit} (black line) is surrounded by a gray shaded region denoting the standard error with $95\%$ confidence intervals on the fit parameters. Note that these curves are the unscaled versions of those plotted in Fig.~2.  The signature of the algorithmic slowdown for \textbf{d)} $d=4$, \textbf{e)} $5$, and \textbf{f)} $6$  is similar to that for in $d=3$ (see Fig.~4). Lines are fits to Eq.~\eqref{eqn:onsetFitG}.}
\label{fig:onset456}
\end{figure*}

\begin{figure}[]
\includegraphics[width=0.9\linewidth]{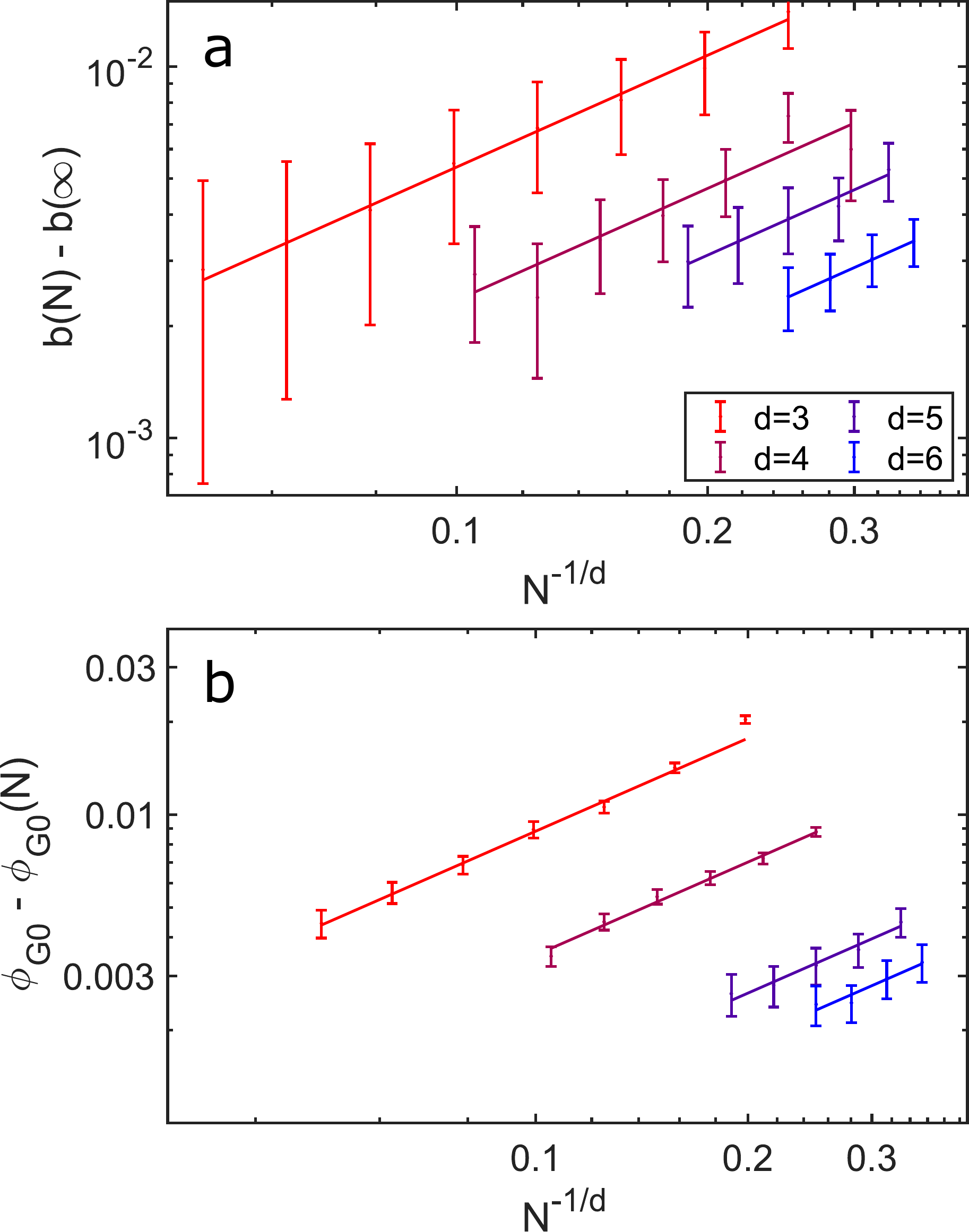}
\caption{\textbf{a)} The scaling of the crossover width fitting parameter $b$ from Eq.~\eqref{eqn:onsetFit} is consistent with ${b(N) - b(\infty) \sim N^{-1/d}}$. \textbf{b)} The scaling of $\phi_\mathrm{G0}$, the zero density limit of $\phi_\mathrm{G}$, from Eq.~\eqref{eqn:onsetFitG}, is consistent with ${\phi_\mathrm{G0} - \phi_\mathrm{G0}(N) \sim N^{-1/d}}$. Curves are offset for visual clarity, and error represent a $95\%$ confidence interval on the fit.}
\label{fig:G0andB}
\end{figure}

The onset curves for both $\phi_\mathrm{J}$ and $\phi_\mathrm{G}$ are given in $d=3$ (Figs.~1 and 4), and the thermodynamic limit is reported for $d=4-6$ (Fig.~2). The intermediate results for ${d=4-6}$ are provided in Fig.~\ref{fig:onset456}.  Equations~\eqref{eqn:onsetFit} and \eqref{eqn:onsetFitG} are each fit simultaneously for all $N$ using standard least-squares methods treating ${\phi_\mathrm{J0}(N)}$ (or ${\phi_{G0}(N)}$), ${\phi_\mathrm{co}(N)}$, and ${b(N)}$ as independent parameters for each curve, and $a$ as a common parameter. The resulting ${\phi_\mathrm{J0}}$, ${\phi_{G0}}$, ${\phi_\mathrm{co}}$, and $b$ are then fitted using the form ${x(\infty) - x(N) = c_1 N^{-\nu/d}}$ with arbitrary constant $c_1$. The exponent $\nu$ is allowed to vary for $\phi_\mathrm{J0}$ and is found to be $\nu \approx 1$ for all $d$ as noted in the text. For the other fitting parameters, we fix $\nu=1$ and verify that the resulting fits are appropriate. Figure~\ref{fig:G0andB} confirms that both $b$ and $\phi_\mathrm{G0}$ scale as $N^{-1/d}$. The resulting fit parameters are given in Table~\ref{tab:fitparams}.

\begin{table}
\begin{center}
\begin{tabular}{c | c | c | c | c | c }
$d$ & $\phi_\mathrm{J0}$ & $\phi_\mathrm{on}$ & $a$ & $b(\infty)$ & $\phi_\mathrm{G0}$ \\
\hline
3 & 0.6487(7) & 0.5101(10) & 0.42(6) & 0.0193(8) & 0.6468(9) \\
4 & 0.4564(18) & 0.346(5) & 0.50(5) & 0.016(4) & 0.4525(6) \\
5 & 0.3083(9) & 0.225(3) & 0.52(7) & 0.009(3) & 0.3015(15) \\
6 & 0.2008(16) & 0.1416(14) & 0.56(5) & 0.0070(9) & 0.195(3) \\
\end{tabular}
\end{center}
\caption{Numerical values for the fits to Eqs. \eqref{eqn:onsetFit} and \eqref{eqn:onsetFitG} in the thermodynamic limit. Errors represent a $95\%$ confidence interval.}
\label{tab:fitparams}
\end{table}


\section{State following finite-size scaling}
\begin{figure}[]
\includegraphics[width=0.9\linewidth]{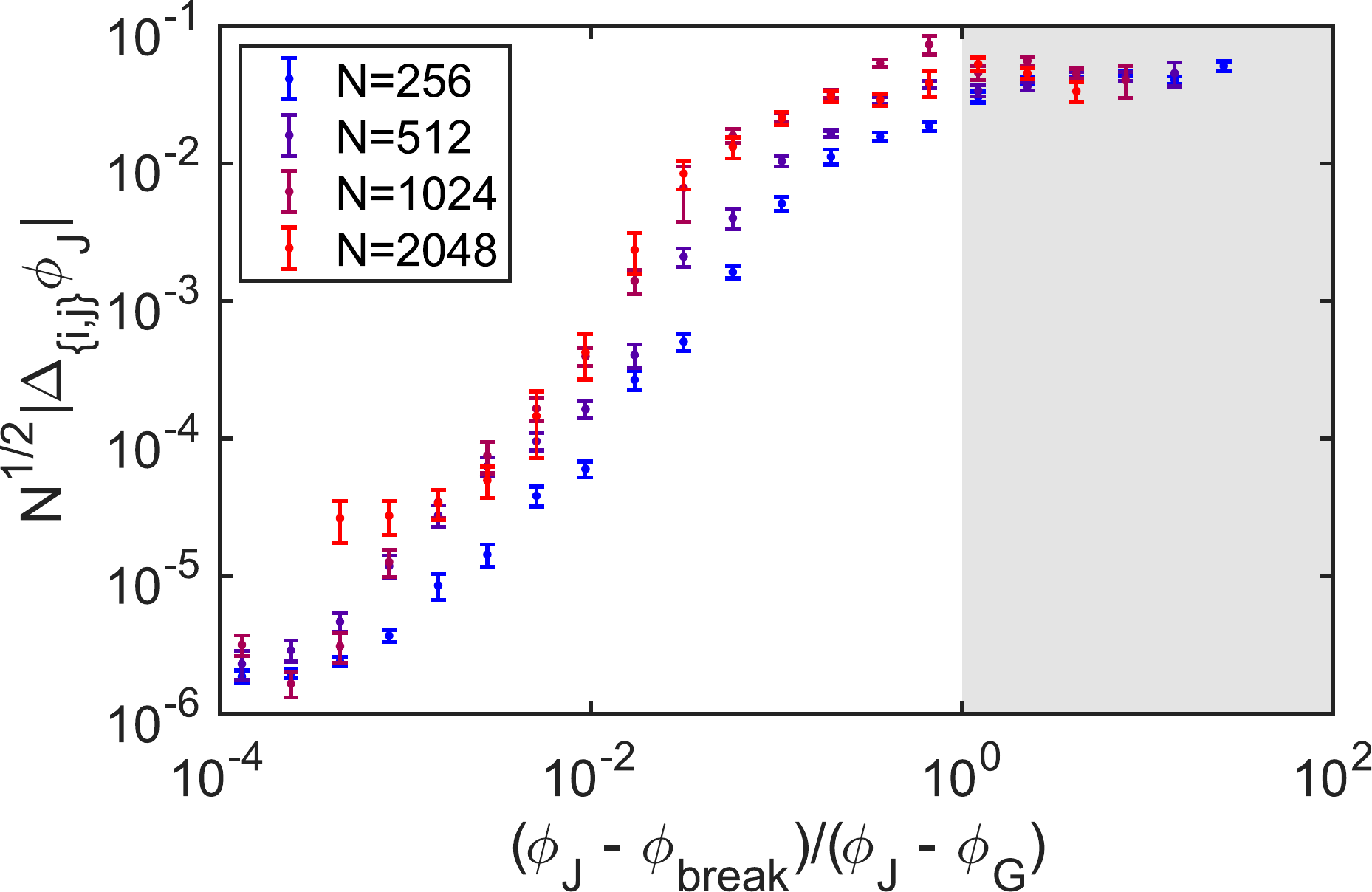}
\caption{Evolution of the jamming density difference,$\Delta_{\{i,j\}} \phi_\mathrm{J}$, between replicas perturbed at $\phi_\mathrm{break}$. Scaling this difference with $N^{-1/2}$ allows a direct comparison the typical density spread between jammed states obtained from a same $\phi_\mathrm{eq}$ (Fig.~\ref{fig:phiJwidth}). Replicas taken at $\phi_\mathrm{break} < \phi_G$ (gray zone) have $\Delta_{\{i,j\}} \phi_\mathrm{J}$ comparable to $\sigma_{\phi_\mathrm{J}}$, but replicas taken at $\phi_\mathrm{break} > \phi_G$ (white zone) converge towards the same jamming density.}
\label{fig:stateFollowFS}
\end{figure}

In Fig.~\ref{fig:gardnerDef}c, differences between replica contact networks are plotted as ${1-(c_i \cap c_j)/(c_i \cup c_j)}$. To show the robustness of the state following routine, we introduce a second metric to test whether replicas tend towards the same state. Equation~\eqref{eqn:phiJfluct} shows that the distance between typical states scales as $\sigma_{\phi_\mathrm{J}} \sim N^{-1/2}$. It is then natural to consider the difference between jamming densities of typical replicas separated at $\phi_\mathrm{break}$, which we denote $N^{1/2} \Delta_{\{i,j\}} \phi_\mathrm{J} $. Figure~\ref{fig:stateFollowFS} shows that systems separated at $\phi_\mathrm{break} < \phi_\mathrm{G}$ have $N^{1/2} \Delta_{\{i,j\}} \phi_\mathrm{J} \sim \sigma_{\phi_\mathrm{J}}$, while systems separated at $\phi_\mathrm{break} > \phi_\mathrm{G}$ begin to converge on the same jammed state.

%

\clearpage
\bibliographystyle{apsrev4-1}
\bibliography{onsetPaper,footnotes}



\end{document}